# Topological magnon bands in a room temperature Kagome magnet

H. Zhang[1], X. Feng[2], T. Heitmann[3], A. I. Kolesnikov[4], M. B. Stone[4], Y. -M. Lu[2], and X. Ke[1]

[1]*Department of Physics and Astronomy, Michigan State University, East Lansing, MI 48824*

[2]*Department of Physics, Ohio State University, Columbus, Ohio 43210, USA*

[3]*Missouri Research Reactor, University of Missouri, Columbia, Missouri 65211, USA*

[4]*Quantum Condensed Matter Division, Oak Ridge National Laboratory, Oak Ridge, Tennessee 37831, USA*

Topological magnon is a vibrant research field gaining more and more attention in the past few years. Among many theoretical proposals and limited experimental studies, ferromagnetic Kagome lattice emerges as one of the most elucidating systems. Here we report neutron scattering studies of $YMn_6Sn_6$, a metallic system consisting of ferromagnetic Kagome planes. This system undergoes a commensurate-to-incommensurate antiferromagnetic phase transition upon cooling with the incommensurability along the out-of-plane direction. We observe magnon band gap opening at the symmetry-protected K points and ascribe this feature to the antisymmetric Dzyaloshinskii–Moriya (DM) interactions. Our observation supports the existence of topological Dirac magnons in both the commensurate collinear and incommensurate coplanar magnetic orders, which is further corroborated by our symmetry analysis. This finding places $YMn_6Sn_6$ as a promising candidate for room-temperature magnon spintronics applications.

Since the experimental discovery of topological insulators [1] predicted by topological band theory [2,3], research on topological materials has experienced rapid growth. While this field was originally focused exclusively on electronic systems, it has quickly diversified into other research arenas. Studies of non-trivial band topology of photon [4], phonon [5] and magnon [6,7] bands are emerging. Topological magnons are important to serve as a potential solution to a long-standing problem in magnon spintronics [8]. Magnon spintronics is considered to be a promising route for circumventing the shortcomings of traditional electronic based devices, since magnons propagate without Joule heating. However, due to the wave nature of magnons, transmitting information (spin wave) unidirectionally is challenging [9]. A natural solution to this problem is offered by leveraging the unique chiral nature of topological magnons. This allows topological magnons to propagate unidirectionally along the edges regardless of the specific device geometry [7].

While topological magnon provides exciting prospects for both fundamental scientific study and potential technological applications [10], the field is rapidly developing and demands more research efforts. In this respect, the theory efforts [11-16] have far outpaced experiments [17-19]. This is partially due to the fact that the exchange interactions in real materials are rarely as simple as what are proposed in theoretical models. However, there is a common thread amidst many proposals, i.e., the quasi-two-dimensional Kagome ferromagnet [11]. In a ferromagnetic Kagome lattice, there are six symmetry-protected band crossings at K-points <1/3 1/3 0>, which retain their degeneracy in the presence of any symmetric exchange interactions (e.g. Heisenberg interactions). The degeneracies at these K-points can only be lifted when anti-symmetric interactions (e.g. Dzyaloshinskii–Moriya interaction) are present, which leads to the opening of magnon gaps [20]. The associated Berry curvature and Chern number of the magnon bands as well as the corresponding edge states have been thoroughly calculated, and it has been shown that these magnon gaps in a Kagome lattice are topological in nature [11,21,22].

To the best of our knowledge, experimental studies of topological magnons have been limited to insulator/semiconductors with low ordering temperatures (below the liquid nitrogen temperature of 77 K) [17-19], which prohibits the development of room temperature magnon spintronic devices utilizing topological magnons. While the quasi-two-dimensional ferromagnetic Kagome lattice is relatively rare in insulating compounds, it is more commonly found in intermetallic materials [23]. For example, the B35 (Strukturbericht Designation) structure type and its derivatives consist of hundreds of intermetallic compounds featuring quasi-two-dimensional Kagome lattices, many of which order magnetically above room temperature. Thus, intermetallic magnets are ideal for the search of functional materials hosting topological magnons. $YMn_6Sn_6$ (YMS), which orders antiferromagnetically around 350 K [24], is one such candidate. There have been some studies of YMS regarding its magnetization [25] and doping effect [26]; however, neutron diffraction study has been limited to polycrystalline samples [27]. Moreover, the magnetic Hamiltonian of YMS is still unknown due to the lack of inelastic neutron scattering (INS) measurements. Previous powder neutron diffraction studies suggest a magnetic ground state consisting of ferromagnetic Kagome sheet of Mn spins [27]. Therefore, YMS is a high temperature magnetic metal potentially hosting topological magnons. Detailed neutron scattering studies of YMS in the single crystalline form is of great interest and importance.

In this Rapid Communication, we present neutron scattering measurements of YMS, an antiferromagnetic metal consisting of ferromagnetic double-layers of Mn Kagome plane. We have observed a commensurate-to-incommensurate antiferromagnetic phase transitions with the incommensurability developing perpendicular to the ab-plane upon cooling. And we have determined the magnetic Hamiltonian by fitting the measured magnon spectra using Linear Spin Wave (LSW) theory. Our experimental results show clear band gap opening at the symmetry-protected K-points (<1/3 1/3 0>). We analyze the spectra by fitting with linear spin wave theory and ascribe the observed band gap opening to the DM interactions with in-plane DM vectors. In conjunction with symmetry analysis, our findings support the existence of non-trivial topological Dirac magnons in YMS.

YMS single crystals were grown using flux method [25]. Magnetic susceptibility of YMS was carried out using a Superconducting Quantum Interference Device (SQUID). Single crystal neutron diffraction measurements were performed on a triple-axis spectrometer (TRIAX) with incident neutron wavelength of 2.359Å (14.7meV) at Research Reactor in University of Missouri. The collimator setting of 60'-40'-S-40'-40' were used in this experiment. The single crystal sample was measured in both (H 0 L) and (0 K L) scattering planes. We also performed INS measurements using a time-of-flight spectrometer (BL-17 SEQUOIA, [28]) at the Spallation Neutron Source (SNS) in Oak Ridge National Laboratory (ORNL) [29]. Multiple pieces of single crystals (total mass ~ 5 g) were coaligned in the (H K 0) scattering plane.

YMS crystallizes in space group P6/mmm (No. 191) with lattice parameters $a = b = 5.536$Å, $c = 9.019$Å, and $\alpha = 90°$, $\beta = 90°$, $\gamma = 120°$ at room temperature. The crystal structure of YMS is illustrated in Figure 1(a). Within each unit cell there are two planes of Mn atoms at $c_1 = 0.25*c$ and $c_2 = 0.75*c$, each of which forms a Kagome web (Fig. 1(b)). Figure 1(c) presents magnetic susceptibility measured with an applied magnetic field of 0.1 T, which clearly shows two magnetic phase transitions occurring at 350 K and 331 K, consistent with previous reports [25]. The magnetic structure associated with these phase transitions remains unclear: an early powder neutron diffraction study of YMS proposed a complex helical spin structure with a non-constant rotation angle as the low temperature magnetic ground state [27].

To resolve the magnetic structure of YMS, we first performed [0 0 L] scan at different temperatures. In Fig. 1(d), we plot the intensity contour map of [0 0 L] peak as a function of temperature and L. One can clearly see that the system undergoes a commensurate-to-incommensurate phase transition upon cooling at $T_{IC}$ = 331K. As shown in the expanded view presented in Fig. 1(e), YMS orders with a commensurate ordering wave vector of (0 0 0.5) at $T_{IC}$ < T < $T_N$ (Note that $T_N$ determined by single crystal neutron diffraction is slightly above 350 K, which unfortunately is beyond the capability of the sample environment used). Below $T_{IC}$, an incommensurate phase with an ordering wave vector of (0 0 0±δ) emerges and coexists with the commensurate phase at 310 K < T < $T_{IC}$; below 310 K the commensurate phase disappears while the incommensurability δ gradually increases with decreasing temperature and eventually stays constant at δ ≈ 0.266 below 50 K. Such a temperature dependent commensurate-to-incommensurate phase transition is an unusual observation, as for most systems reported in literature an incommensurate structure gradually 'locks-in' to a commensurate one when temperature decreases [30].

In order to determine the magnetic structure of YMS, we have collected a series of nuclear and magnetic Bragg peaks at T = 350 K and T = 6 K and performed Rietveld refinement [31]. At both temperatures, we found that each Kagome double layer of Mn moments within one nuclear unit cell is aligned ferromagnetically with spins pointing in the ab plane as illustrated in Fig. 1(a). This indicates strong ferromagnetic in-plane and interplane interactions within each double layer. Thus, each double layer can be treated as a single unit. Between such neighboring units along the c-axis, the spins are rotated by an angle $\theta$. At T = 350 K, the angle $\theta$ = 180° and the magnetic structure is commensurate with the propagation vector $\vec{k} = (0\ 0\ 0.5)$. At 350 K, the ordered moment size is ≈ 0.22(2) $\mu_B$/Mn. As temperature decreases, so does $\theta$, and the spin structure becomes helical with an incommensurate propagation vector. The magnetic structure at $T = 6$ K is described by the propagation vector $\vec{k} = (0\ 0\ 0.266)$, which correspond to a spin rotational angle of $\theta = 95.6°$. At 6 K, the ordered moment size is ≈ 2.20(1) $\mu_B$/Mn.

As discussed in the introduction, the ferromagnetic Kagome plane of Mn moment makes YMS a candidate for hosting topological magnon bands. In order to determine the nature of magnon bands and the underlying magnetic exchange interactions of this system, we have performed time-of-flight INS measurements. Large sets of 4D data containing neutron scattering intensity with both energy and momentum transfer ($I(\vec{q}, \Delta E)$) are obtained, which can be used to infer the spin dynamics of systems.

In Fig. 2 we present the intensity contour maps $I \cdot \Delta E\ (q_x, q_y)$ measured at $T = 5$ K and their corresponding simulations. The data are presented in Cartesian coordinates to illustrate the hexagonal symmetry of the spectra. The x-axis is along the (H 0 0) direction and the y-axis is along the (-H 2K 0) direction. Blue dashed lines represent the boundary of Brillouin zones. Figure 2 (a) plots the contour map measured at $\Delta E = 55.8$ meV with L centered at 0 (L = [-0.30 0.30]), in which a hexagon-like excitation can be clearly seen. This excitation is attributed to the magnon modes propagating in the Kagome plane. The spectra of these magnon modes along $\Gamma$-$K_4$-$M_5$-$\Gamma$ are presented in Fig. 3(a). This dispersion can be fitted using the observed magnetic ground state and a dominating ferromagnetic nearest-neighbor interaction ($J_1$ denoted in Fig. 1(b)) as an initial model. Upon closer examination of the low energy spectrum in Fig. S2(b) [22], one can see that there is a slight asymmetry between the left and right branches around $\Gamma$ point. This is due to the easy-plane anisotropy of this system. This anisotropy agrees with the fact that the magnetic ground state consists of spins oriented in the ab-plane. The weak dispersion between $K_4$ and $M_5$ indicates that further neighbor interactions are weak, but not negligible.

To address the interplane coupling in the system, we first plot the intensity map $I \cdot \Delta E\,(q_x, q_y)$ with L centered at 0.5 (L = [0.35 0.65]) and $\Delta E$ = 30 meV, as shown in Fig. 2(b). Within each Brillouin zone, in addition to the ring-like feature, a new feature with high intensity at the zone center ($\Gamma$) is clearly observed. The dispersion spectra along the $K_4$-$K_1$-$K_6$ directions are shown in Fig. 3(b) (inset is the trajectory). Above 30 meV, a double ring feature is clearly seen, which arises from strong intra-bilayer coupling ($J_3$).

To better understand the experimental data and extract the magnetic exchange interactions, we construct a model Hamiltonian shown in Eq. (1) and use linear spin wave (LSW) theory to compare to the observed spectra [32]. Note that we take S = 1 considering the measured moment of ~ 2.20(1) $\mu_B$/Mn discussed previously. We consider the first three nearest neighbor interactions ($J_1$, $J_4$, $J_7$) in each Kagome plane, as illustrated in Fig. 1(b). Note that there is no atom at the center of each hexagon, thus the interaction between Mn atoms along the diagonal of each hexagon ($J_8$) is expected to be very weak due to the large Mn-Mn distance. Between the Kagome planes shown in Fig. 1(a), we consider the intra-bilayer couplings ($J_3$, $J_6$, $J_{10}$) and inter-bilayer couplings ($J_2$, $J_5$, $J_9$). As inferred from the low energy spectra (Fig. S1(b)), an easy-plane single ion anisotropy needs to be included in the model as well.

$$H_1 = \sum_{<i,j>} J_{ij} \left(\vec{S}_i \cdot \vec{S}_j\right) + A \sum_{<i>} \left(S_{i,x}^2 + S_{i,y}^2\right) \quad (1)$$

The calculated dispersion curves and the associated spectra along various trajectories, together with their comparison with experimental results, are shown in Fig. 2 and Fig. 3. Overall, the calculated results are in good agreements with the experimental data. The magnon spectra observed in the experiment is broader than what is anticipated after considering instrument resolution effect. This is presumably due to magnon-itinerant electron interactions (Stoner excitation) in metallic ferromagnets [33]. The extracted exchange parameters $J$s and the single ion anisotropy term $A$ are listed in Table 1.

The details of the fitting process using LSW theory are described in the Supplemental Materials [22]. Several major features are worth pointing out. First, although $J_1$ is the dominating exchange interaction in the Kagome plane, it is necessary to include the second ($J_4$) and third nearest-neighbor interactions ($J_7$) to get better fit to the spectra in Fig. 3. Second, the ferromagnetic intra-bilayer coupling ($J_3$) breaks the double degeneracy of each band by pushing them apart in energy (see red dashed arrows in Fig. 3(a, b)). The intensity of the split band is L dependent. In Fig. 3(a) with L centered at 0 (L = [-0.30 0.30]) the upper branch shows no intensity, while in Fig. 3(b) with L centered at 0.5 (L = [0.35 0.65]) both branches are clearly visible. Third, the inter-bilayer couplings ($J_2$, $J_5$, $J_9$) cannot be determined in this experiment. This is because of the nearly perpendicular spin alignment (95.6°) between neighboring bilayers along the c-axis, which is associated with the incommensurate spin structure at $T = 6$ K, such that $\vec{S}_i \cdot \vec{S}_j$ is close to zero.

Having determined the leading exchange interactions in the magnetic Hamiltonian, we will next discuss DM interaction and its role on the topology of magnon bands. Figure 4(a) present the magnon spectrum measured along the $K_1$ - $K_2$ direction. Interestingly, magnon band gap opening is clearly observable around K points. To better quantify the size of this magnon gap, we performed constant Q-cut and plotted the I· ΔE *vs* E curve in Fig. 4(b) by integrating the intensity along the $K_1$, $K_2$ and $K_5$. This data can be nicely fitted with two Gaussian peaks centering at 56.3 (0.9) meV and 78.5 (1.4) meV and with a magnon gap centering around 68(2) meV. The band gap opening at these K points cannot be captured using $H_1$ only in the spin Hamiltonian; instead DM interaction needs to be taken into account, as presented in equation (2). Considering that Mn spins point in the ab-plane, the DM interaction with in-plane DM vectors [34] illustrated in Fig. 4(c) is considered. Note that the addition of DM interactions has little effect on the previous fitting results shown in Fig. 2 and Fig. 3. The calculated dispersion curves using LSW theory [32] are superimposed with the measured spectra shown in Fig. 4(a).

$$H = H_1 + \sum_{<i,j>} (\vec{D}_{ij} \cdot (\vec{S}_i \times \vec{S}_j)) \quad (2)$$

One can clearly see good overall agreement between the experimental data and theoretical calculations. The fitting exchange parameters $J$s are the same as what are described above and listed in Table 1, and the D value is 2.4(0.4) meV. It is known that the DM interaction gives rise to complex hopping matrix elements in the Hamiltonian and acts as a vector potential, which may open topologically non-trivial band gaps in magnon systems [17,19]. Thus, our observation of magnon band gap opening ascribed to an in-plane DM interaction opens the possibility of identifying topological magnons in the system.

To address the topological nature of the magnon bands in YMn$_6$Sn$_6$, we first proceed with symmetry analysis [22] for the high temperature commensurate magnetic phase with $\vec{k} = (0\ 0\ 0.5)$. Considering only Heisenberg interactions in the system, the LSW Hamiltonian is invariant under the following symmetry operation: $G_1 = U(1)_{S^x} \rtimes Z_2^{\tilde{C}_{2,z}}, \tilde{C}_{2,z} = \mathcal{T} \cdot e^{i\pi \hat{S}^z} \cdot C_{2,z}$. Here, $U(1)_{S^x}$ refers to the spin rotational symmetry; $\tilde{C}_{2,z}$ is the combination of time-reversal symmetry and two-fold rotational symmetry in the z-direction. In the commensurate phase, if we neglect the easy-plane anisotropy, symmetry G$_1$ protects a Dirac-type band touching at K(K') in the same way as in graphene, featuring a quantized $\pi$-Berry phase for closed contour around the band touching points. To verify this, we carried out LSW simulation using a minimal $J_1$-$J_2$-$J_3$-DM model with $J_1$ = -28.75 meV, $J_2$ = 5 meV, and $J_3$ = -23.75 meV. The calculated dispersion is shown in Fig. 4(d) where one can clearly observe the expected Dirac-type band touching at K(K'). Our numerical computation confirms the $\pi$-Berry phase around these band touching points. Such topological Dirac magnons are pinned at K(K') by the 3-fold crystal rotational symmetry along z-direction. Introducing DM interactions breaks both $U(1)_{S^x}$ symmetry and 3-fold rotation, leading to the band gap opening at K(K') that we observed at low temperature ($T$ = 5 K, $\vec{k}$ = (0 0 0.266)). However, these Dirac magnons remain robust and protected by the remaining $\tilde{C}_{2,z}$ symmetry and are relocated along the $\Gamma$-K(K') direction. They are featured by a topological $Z_2$-index known as the 1$^{st}$ Stiefel-Whitney [35,36]. These bulk Dirac magnons also give rise to localized topological surface magnons, similar to the protected zigzag edge states of graphene [37].

While protected in collinear and coplanar magnetic orders with $\tilde{C}_{2,z}$ symmetry, any small out-of-plane magnetization (e.g. induced by an out-of-plane field) will break the $\tilde{C}_{2,z}$ symmetry and gap out the topological Dirac magnons on the $\Gamma$-K(K') line, leading to a direct gap at every k between the two lower magnon bands. In the commensurate collinear order, there is an inversion symmetry $I$ centered between the two layers within each ferromagnetic bilayer. Using the inversion symmetry indicator for Chern numbers [38], we can show that once the Dirac magnons are gapped out, the first and second lowest magnon bands have a non-zero and odd Chern number, indicating nontrivial topology of these magnon bands and associated chiral surface magnons [39].

In summary, YMn$_6$Sn$_6$ hosts ferromagnetic Kagome bilayers which antiferromagnetically couple with neighboring bilayers. We have found that this material exhibits an uncommon commensurate-to-incommensurate magnetic phase transition upon cooling with the incommensurate direction along the c axis. We further show that the magnon bands feature band gap opening at K(K') points, which are ascribed to the asymmetric DM interaction with in-plane DM vectors. Our symmetry analysis suggests non-trivial topological nature of the magnon bands in this system. This study, combined with the very recent discovery of large anomalous topological Hall effect [40], places YMn$_6$Sn$_6$ as a rare example in which both electron and magnon bands are topologically non-trivial.

The work at Michigan State University was supported by the National Science Foundation under Award No. DMR-1608752. The work at Ohio State University was supported by National Science Foundation under Award No. DMR-1653769. A portion of this research used resources at the Spallation Neutron Source, a DOE Office of Science User Facility operated by the Oak Ridge National Laboratory.

**Table 1**. Exchange integrals ($J$s), in-plane anisotropy (A), and DM interaction (D) obtained from the LSW fitting to measured magnon spectra. The inter-bilayer interactions, $J_2$, $J_5$ and $J_9$, cannot be determined in this experiment, as discussed in main text. The numbers inside the parenthesis are error bars.

| Bond | $J_1$ | $J_2$ | $J_3$ | $J_4$ | $J_5$ | $J_6$ | $J_7$ | $J_8$ | $J_9$ | $J_{10}$ | A | D |
|---|---|---|---|---|---|---|---|---|---|---|---|---|
| Strength (meV) | -28.75 (0.75) | \ | -23.75 (0.75) | -5.25 (0.25) | \ | 0 (0.25) | 6.25 (0.25) | 0 (0.25) | \ | 2.5 (0.25) | -5.0 (1) | 2.4 (0.4) |

**Figure Captions:**

Figure 1. (a) Crystal and magnetic structure of YMn$_6$Sn$_6$. Yttrium (Y) atoms are represented by green spheres, Tin (Sn) atoms are represented by grey spheres. Manganese (Mn) atoms are represented by purple spheres while their moments are represented by purple arrows. The Mn atoms form a Kagome structure within the ab-plane, as highlighted by the light blue plane. The magnetic structure of YMn$_6$Sn$_6$ at $T = 350$ K (propagation vector $\vec{k} = (0\ 0\ 0.5)$) consists of bilayers of ferromagnetic Kagome sheets ($\vec{m}//\vec{a}$) alternating along the c-axis. Intra-bilayer and inter-bilayer interactions are labelled. (b) A top view of the Kagome layer of Mn atoms. The in-plane 1$^{st}$ ($J_1$), 2$^{nd}$ ($J_4$) and 3$^{rd}$ ($J_7$) nearest neighbor interaction are represented by black, blue and red bonds respectively. (c) Magnetic susceptibility measured with 0.1T magnetic field applied along a-axis. Inset shows an expanded view of the high temperature region. (d, e) [0 0 L] peak intensity as a function of temperature and L.

Figure 2. (a) Inelastic neutron scattering (I· ΔE, E$_i$ = 120 meV) intensity map at $\Delta E = 55.8\ meV$ for L = [-0.3 0.3]. (b) Inelastic neutron scattering (I· ΔE, E$_i$ = 60 meV) intensity map at $\Delta E = 30\ meV$ for L = [0.35 0.65]. (c-d) The corresponding intensity map simulated using the model described in the main text.

Figure 3. (a) Inelastic neutron scattering (I· ΔE) spectrum observed along Γ(2 0 0)-K$_4$(5/3 -1/3 0)-M$_5$(2 -1/2 0)- Γ(2 0 0) experiment (E$_i$ = 120 meV and 60 meV). Inset shows the trajectory in the reciprocal space. The black curve is the simulated dispersion obtained using SpinW [32]. (b) Spectra through H$_4$(5/3 -1/3 1/2)-H$_1$(7/3 1/3 1/2)-H$_6$(8/3 -1/3 1/2) along [110] and [1$\bar{2}$0] with L center at 0.5 (L = [0.35 0.65]) observed in the experiment. A double cone intensity map is observed as a result of the inter-plane coupling within the ferromagnetic Kagome bilayers. (c) Spectra

through K$_3$(4/3 1/3 0)-K$_1$(7/3 1/3 0)-K$_6$(8/3 -1/3 0)-K$_3$(4/3 1/3 0) along [100], [1$\bar{2}$0] and [$\bar{2}$10] observed in the experiment. The trajectory is shown in inset.

Figure 4. (a) Inelastic Neutron scattering spectra (I· ΔE) through K1(1/3 7/3 0)-K2(-1/3 8/3 0); (b) I· $\Delta E - \Delta E$ plot after integrating the data from K1(1/3 7/3 0), K2(-1/3 8/3 0) and K5(7/3 -8/3 0). Both experimental data and fitting curve described in the main text are shown. (c) An illustration of the in-plane DM interaction in the magnetic Hamiltonian; (d) Linear Spin Wave simulation along Qy (Qx = 0.5, Qz = 0) results for the general $J_1$-$J_2$-$J_3$-DM model. Dirac band touching points are circled out in red color.

Figure 1.

H. Zhang et al,

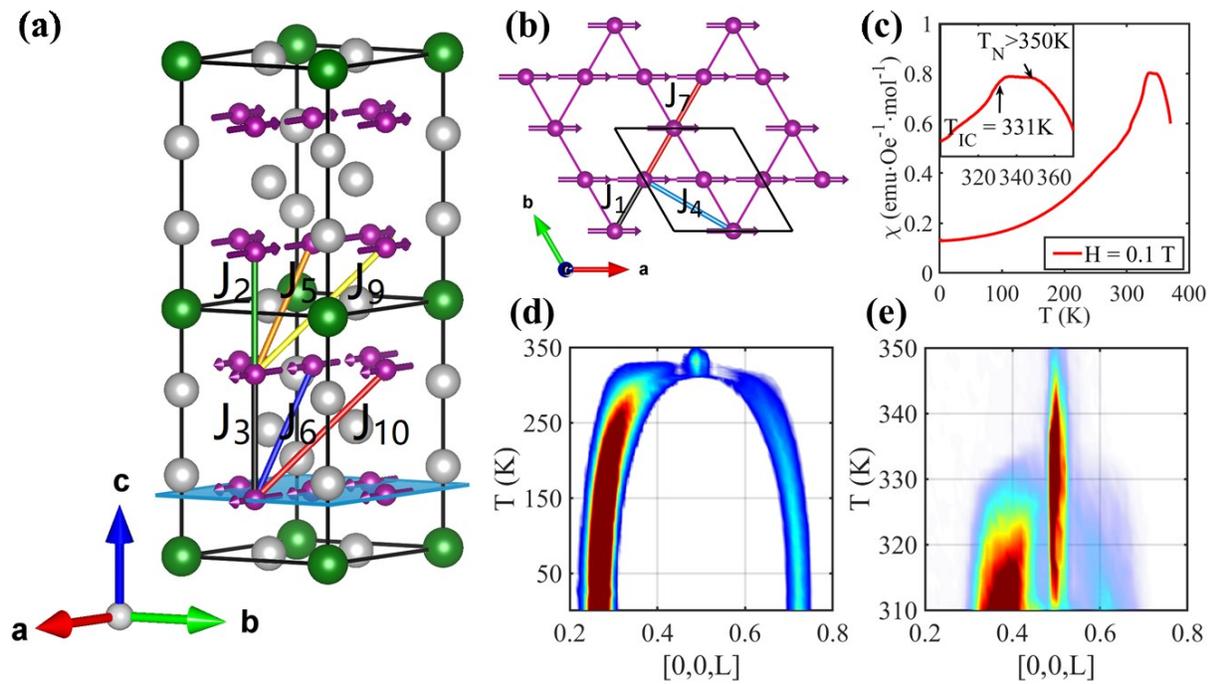

Figure 2.

H. Zhang et al,

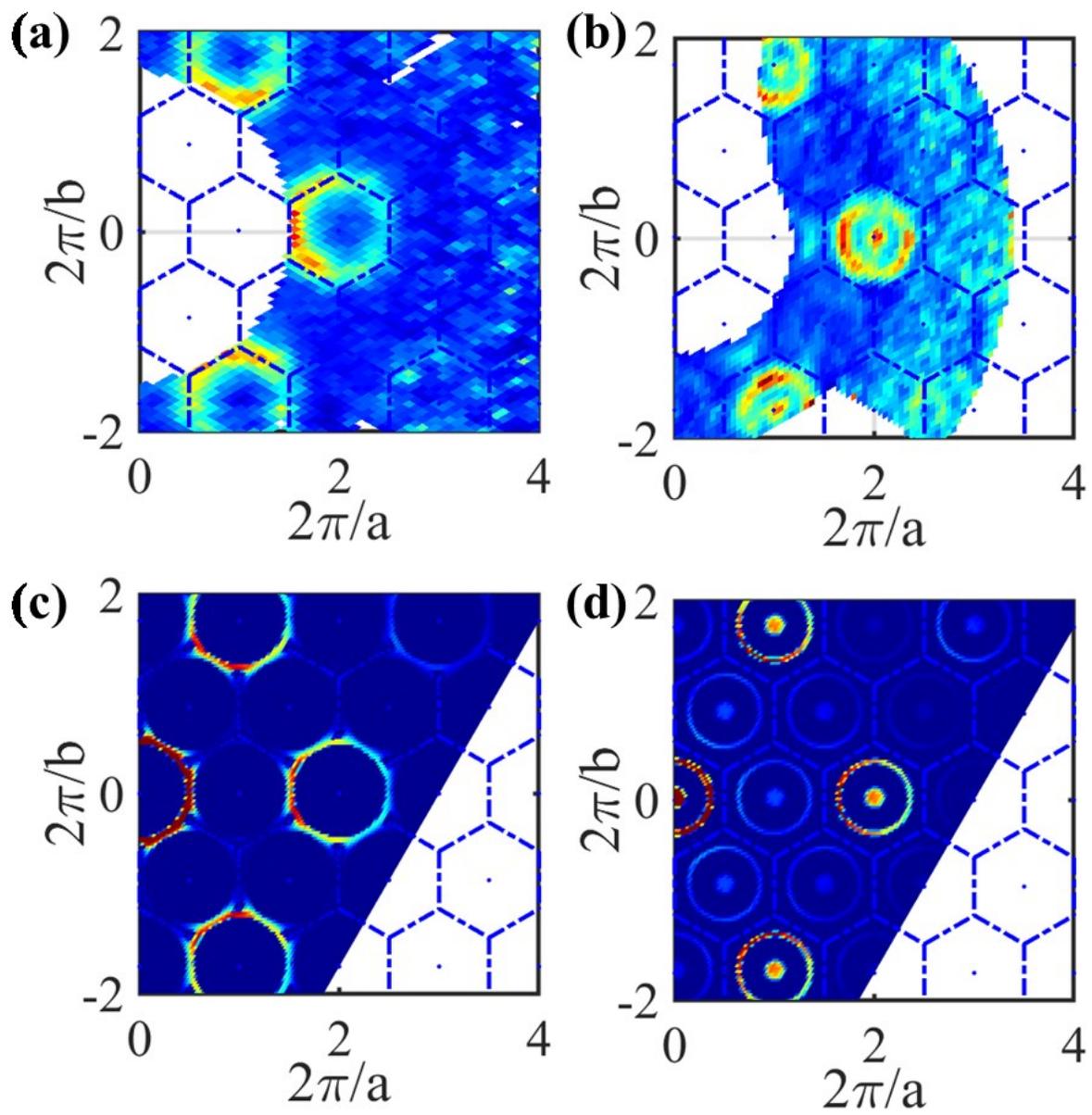

Figure 3.

H. Zhang et al,

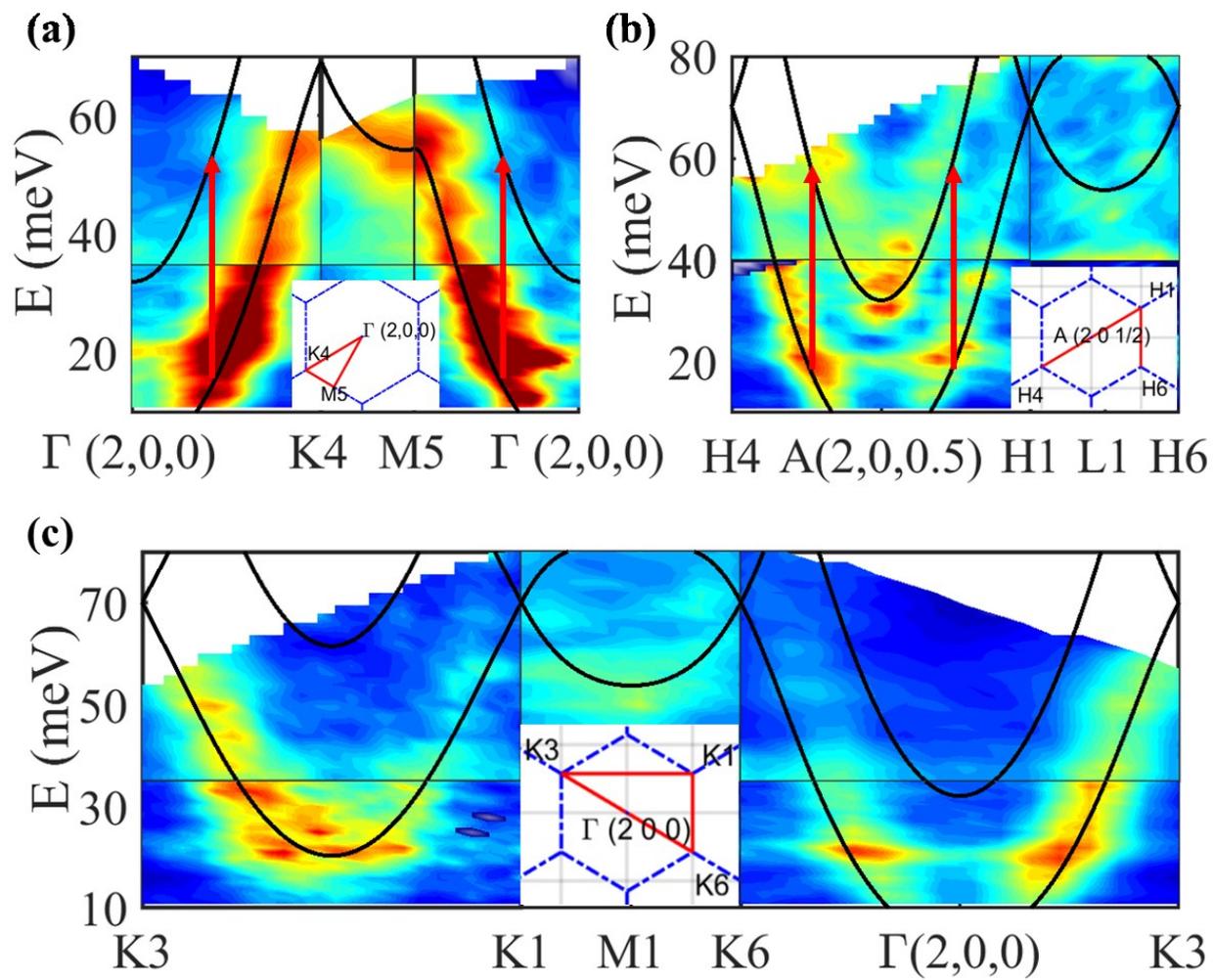

Figure 4.

H. Zhang et al,

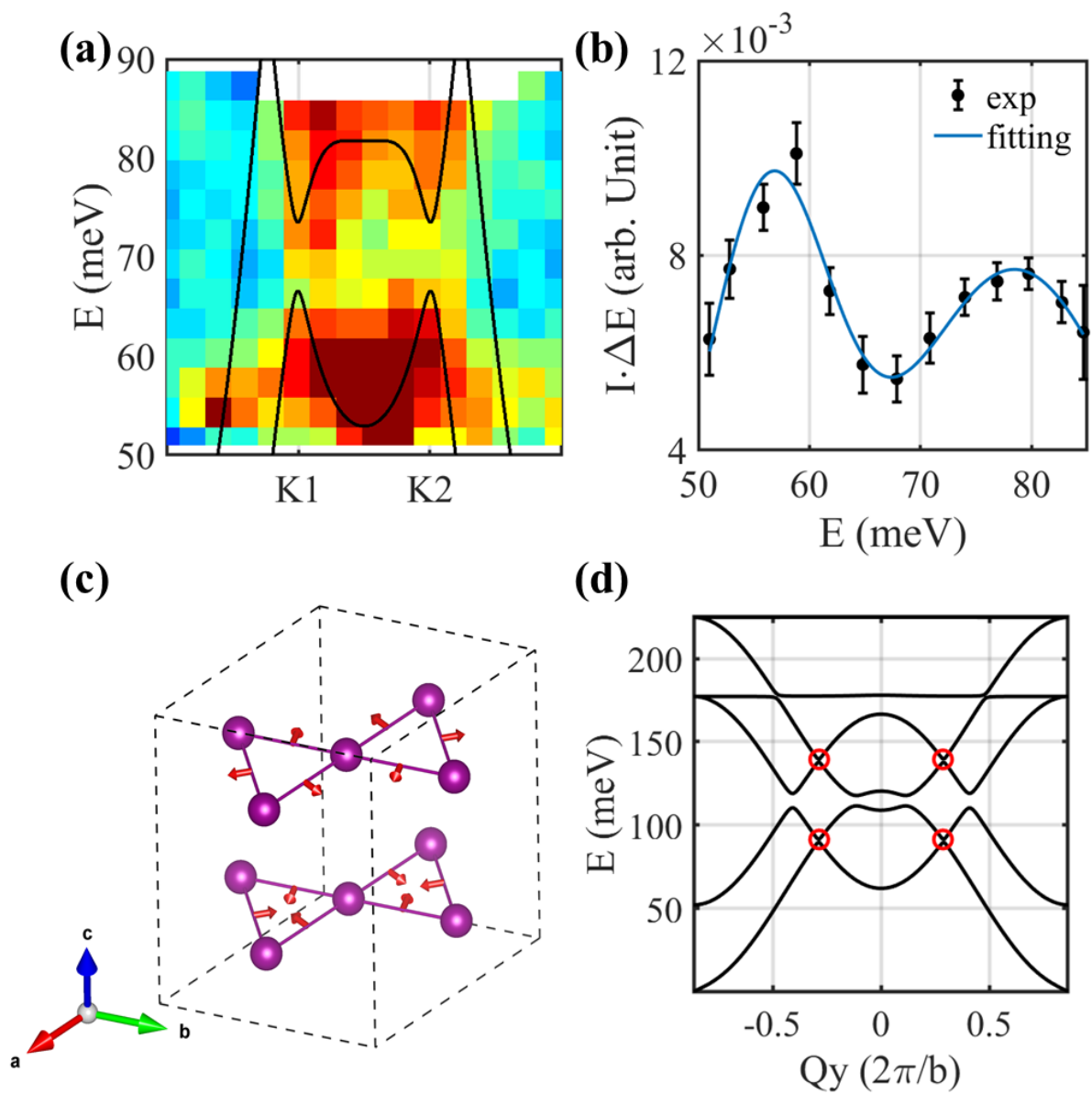